\documentclass[aip,graphicx,amsmath,amssymb,reprint]{revtex4-1}
\usepackage{graphicx}
\usepackage{dcolumn}
\usepackage{bm}

\usepackage[utf8]{inputenc}
\usepackage[T1]{fontenc}
\usepackage{color}
\usepackage[normalem]{ulem}

\begin{document}

\preprint{AIP/123-QED}

\title{A perspective on scaling up quantum computation with molecular spins}

\author{S. Carretta}
\affiliation{ 
Department of Mathematical, Physical and Computer Sciences, University of Parma (Italy) 
}
\affiliation{UdR Parma, INSTM, I-43124 Parma, Italy}
\author{D. Zueco}%
\affiliation{ 
Instituto de Nanociencia y Materiales de Arag\'on (INMA), CSIC-Universidad de Zaragoza, 50009 Zaragoza (Spain)
}%
\author{A. Chiesa}
\affiliation{ 
Department of Mathematical, Physical and Computer Sciences, University of Parma (Italy) 
}%
\affiliation{UdR Parma, INSTM, I-43124 Parma, Italy}
\author{\'{A}. G\'{o}mez-Le\'{o}n}
\affiliation{ 
Instituto de F\'{\i}sica Fundamental, CSIC, 28006 Madrid (Spain)
}%
\author{F. Luis}
\affiliation{ 
Instituto de Nanociencia y Materiales de Arag\'on (INMA), CSIC-Universidad de Zaragoza, 50009 Zaragoza (Spain)
}%


\date{\today}

\begin{abstract}
Artificial magnetic molecules can contribute to progressing towards large scale quantum computation by: a) integrating multiple quantum resources and b) reducing the computational costs of some applications. Chemical design, guided by theoretical proposals, allows embedding nontrivial quantum functionalities in each molecular unit, which then acts as a microscopic quantum processor able to encode error protected logical qubits or to implement quantum simulations. Scaling up even further requires "wiring-up" multiple molecules. We discuss how to achieve this goal by the coupling to on-chip superconducting resonators. The potential advantages of this hybrid approach and the challenges that still lay ahead are critically reviewed. 
\end{abstract}

\maketitle

%

\section{\label{Intro:section} Introduction}

A crucial challenge for the development of quantum technologies is to reach a
computational power able to tackle problems of social and economical
value.\cite{Gibney2014,Mohseni2017} Estimating what is necessary depends on details
of the problem itself and of the platform chosen to solve it. Yet, it appears that
performing quantum simulations or prime-number factorization of relevance to
applications will demand operating over many thousands, even millions of
qubits.\cite{Fowler2012} This daunting prediction arises not only from the
complexity of such problems, but also from the need of protecting quantum
operations from noise and the fact that quantum error correction (QEC) is based on
increasing the number of physical qubits encoding each logical (error-protected) qubit.\cite{Knill1997,Terhal2015}

Although there is hope that the already available Noisy Intermediate-Size Quantum devices (NISQs) will be useful for some specific
tasks,\cite{Preskill2018,Martinis2019,Cross2019} it makes sense to consider
alternatives. Electron spins in semiconductor quantum dots\cite{Loss1998} or atomic impurities\cite{Pla2012} represent natural candidates to attain high levels of integration. Progress along this direction, although encouraging,\cite{Morton2011,Awschalom2013,Hendrickx2021} faces problems of qubit reproducibility similar to those encountered with other circuits fabricated by top-down lithography or of qubit tunability. 

\begin{figure}
\includegraphics[width = 0.9\columnwidth]{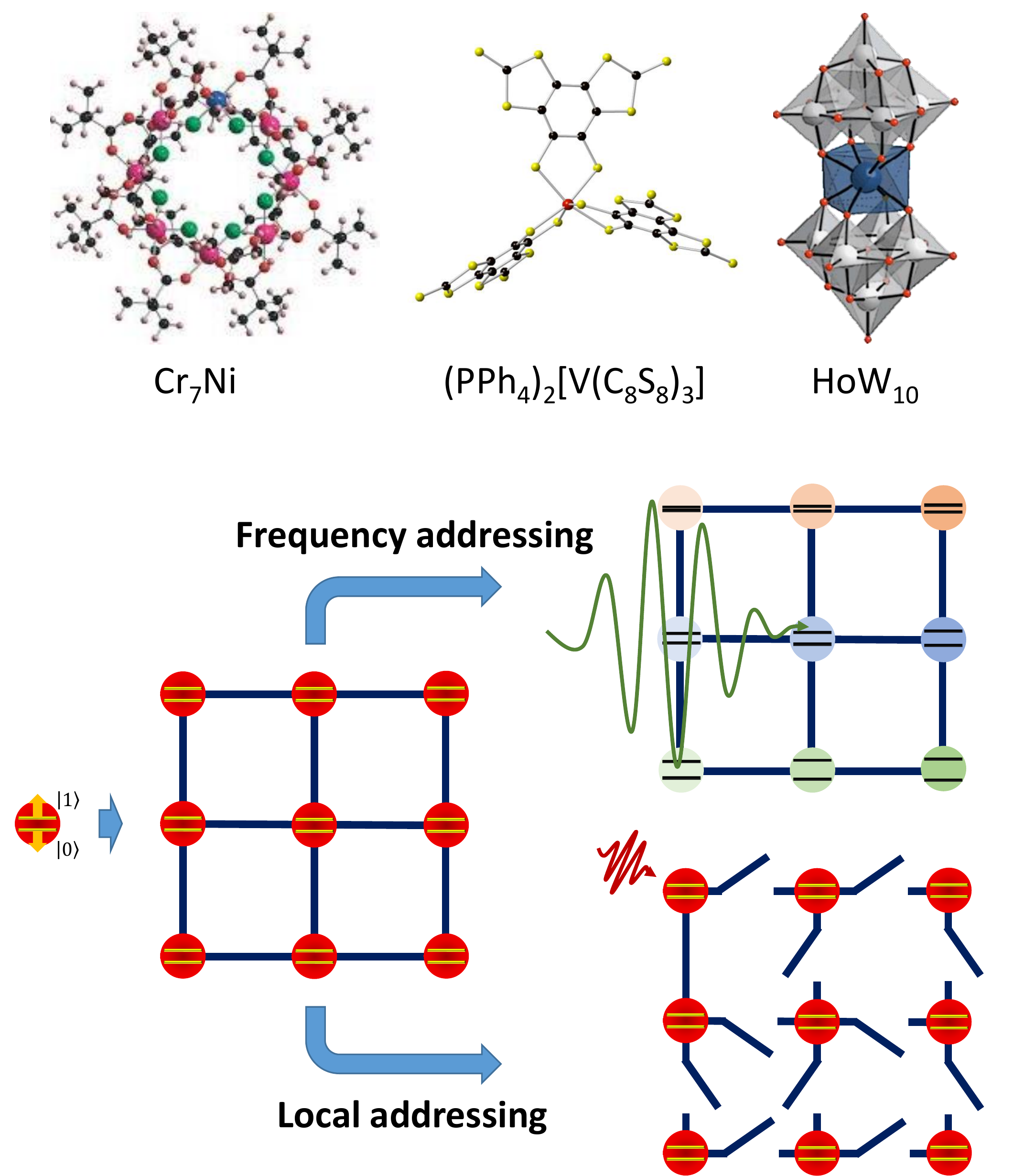}
\caption{\label{Molecular-scaling:figure} (a) Molecular structures of three molecular spin qubits. In the first two,\cite{Ardavan2007,Zadrozny2015a} the qubit is encoded in the states of a spin $S=1/2$. For HoW$_{10}$,\cite{Shiddiq2016} it is defined by the tunnel split states arising from the $m = \pm 4$ total angular momentum projections. (b) Illustration of scaling up by a global frequency addressing and by a local addressing with switchable qubit-qubit interactions.}
\end{figure}

Here, we focus on a different class of spin systems, based on artificial magnetic
molecules.\cite{Gaita2019,Atzori2019} They combine a microscopic, thus close to perfectly reproducible, nature with the ability of chemically designing their
properties. Each of them consists of one to a few magnetic ions stabilized and protected by a shell of organic ligand molecules (Fig.
\ref{Molecular-scaling:figure}). Molecules with an effective $S=1/2$ ground 
state provide the simplest qubit realizations but, as it is discussed below, there exist many other appealing possibilities. 

Our aim is to discuss the potential that such molecular building blocks have to progress towards large-scale quantum computation and the advantages they offer for the implementation of some specific applications. We consider two alternatives for scaling-up, which are schematically illustrated by Fig. \ref{Molecular-scaling:figure}. The first is based on the asymmetry between different qubits in the array (e.g. each having a different frequency) and on the interactions between them. The ensuing energy level anharmonicity then allows one to address each operation by simply choosing the adequate frequency (or "color") of a resonant electromagnetic pulse acting on the whole array. This strategy allows scaling up by "Chemistry", i.e. within each molecule. The second option involves a local control over each qubit and over its interactions with the rest. It relies on the very challenging goal of controlling and "wiring up" individual molecular spins.

\section{\label{Moldesign:section} Scaling up within each molecule: molecular quantum processors}

A characteristic trait of molecular systems is the nearly boundless opportunities to tune their physical properties via changes in composition and structure. The molecular design
allows, for instance, suppressing decoherence by either removing "magnetic noise" sources
\cite{Ardavan2007,Wedge2012,Martinez-Perez2012,Bader2014,Zadrozny2015a,Yu2016,Atzori2016b}
(e.g. replacing some of the molecules in the crystal by non-magnetic derivatives, dissolving
them in adequate solvents or reducing the number of nuclear spins) or by encoding the qubit
states in "decoherence-free" subspaces formed near level
anticrossings.\cite{Shiddiq2016,Zadrozny2017,Collet2019,Rubin2021} The application of these
methods has led to very significant improvements in spin coherence times $T_{2}$ which for
some examples are near one ms.\cite{Zadrozny2015a} 

Chemical design can also be exploited to expand the available computational space from single qubits to $d-$dimensional qudits at the level of a microscopic physical object. An option is to create molecular structures hosting several magnetic centres.\cite{Aromi2012} Examples include molecular dimers and trimers of lanthanide ions,\cite{Luis2011,Aguila2014,Macaluso2020} as well as supramolecular structures able to bind several well-known molecular qubits, such as [Cr$_{7}$Ni], and combine them with other $S=1/2$ complexes.\cite{Ardavan2015,Ferrando2016a,Fernandez2016}

\begin{figure}
\includegraphics[width = 0.9\columnwidth]{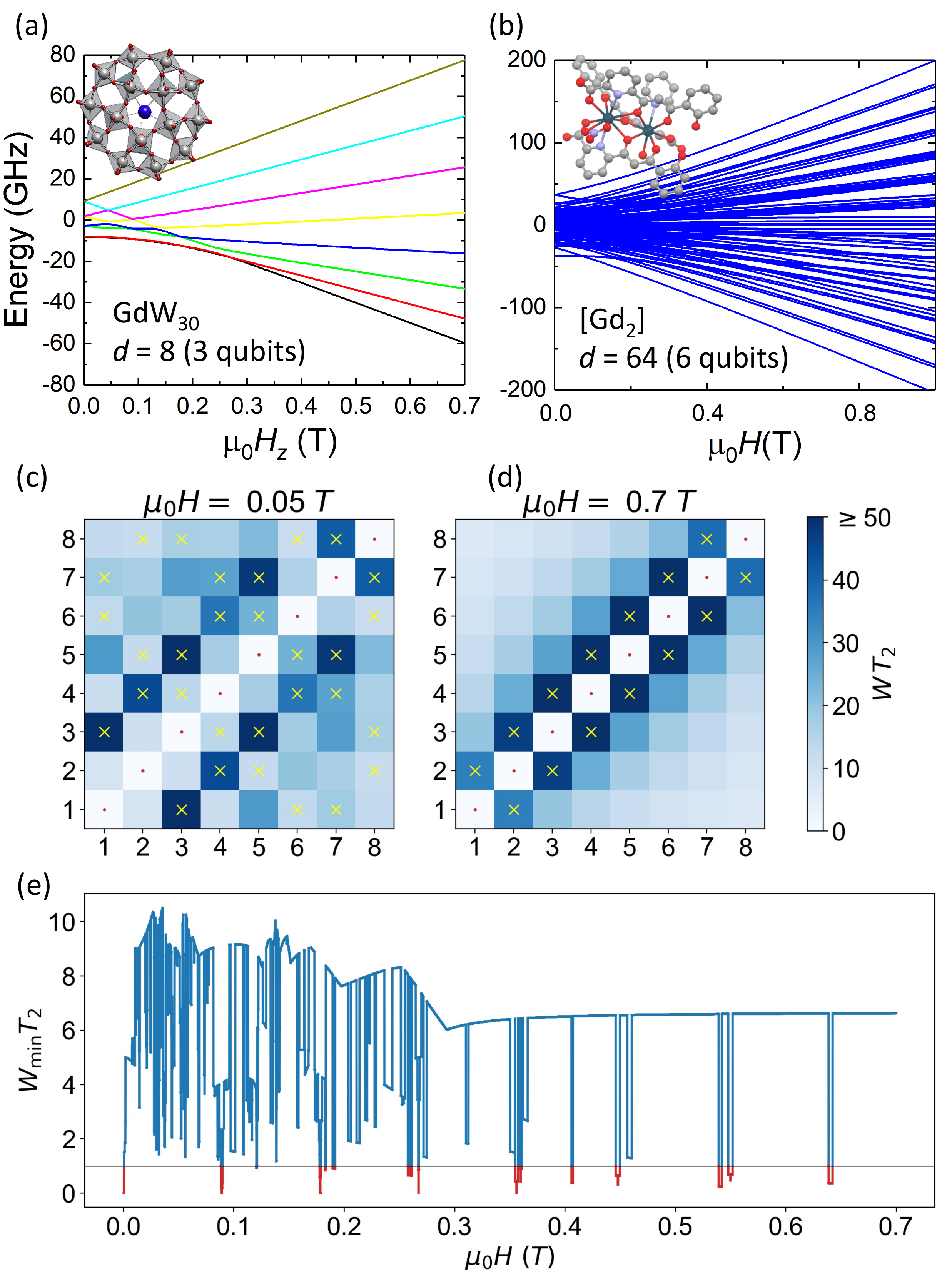}
\caption{\label{Universality:figure} Universality conditions for molecular spin qudits. (a) and (b), energy levels of two molecular spin qudits based on the $S = 7/2$ states of Gd$^{3+}$ ions: GdW$_{30}$ with dimension $2^{3} = 8$ and [Gd$_{2}$] with dimension
$2^{6} = 64$.\cite{Jenkins2017,Luis2020} Their structures are shown as insets. (c) and (d), rates $W_{n,m}$ of quantum operations, performed by sequences of resonant electromagnetic pulses, linking basis states $n$ and $m$ of GdW$_{30}$ at two different magnetic fields. The yellow crosses mark the direct resonant transitions that have a Rabi frequency $\Omega_{\rm R} > 1/T_{2}$, with $T_{2} \simeq 1 \mu$s the spin  coherence time. The red dots mark the trivial identity operation. (e) Universality parameter 
$W_{\rm min}T_{2}$ for GdW$_{30}$ as a function of magnetic field, where $W_{\rm min}$ is the smallest operation rate between any pair of basis states. The red traces signal magnetic 
fields where the system becomes non-universal as a result of accidental degeneracies between two non-forbidden resonant transitions.}
\end{figure}

An alternative is to exploit internal spin degrees of freedom. For instance, the electronic spin $S=7/2$ of a Gd$^{3+}$ ion defines $2^{3} = 8$ discrete levels. In a well-chosen
molecular coordination (see Fig. \ref{Universality:figure}), leading to a sufficiently weak
magnetic anisotropy and correspondingly small level splittings, these states can encode a 
$d = 8$ qudit or $3$ qubits.\cite{Jenkins2017} It is also possible to make use of the metal ions' nuclear
spin states.\cite{Moreno-Pineda2017,Moreno-Pineda2018} The hyperfine
coupling to the electronic spin splits these levels and considerably
speeds up the rates at which such states can be coherently manipulated by electromagnetic
pulses.\cite{Thiele2014,Hussain2018,Gimeno2021} The different strategies can also be combined
to further increase the qudit dimension. For instance, molecular structures with several
magnetic ions, each acting as a qudit, can be synthesized. An illustrative example of a Gd
dimer ($2^6=64$ levels or $6$ qubits) is shown in Fig.
\ref{Universality:figure}.\cite{Luis2020}

The crucial question is then whether one of these molecular qudits is able to
 implement {\em any} quantum algorithm.\cite{Preskill2018,DiVincenzo2000} Even though the physical operation principles are quite similar, magnetic molecules have a crucial 
advantage over NMR quantum computing with organic molecules.
\cite{Vandersypen2004} The sizeable energy splitting between the ground and excited levels, larger than $400$ MHz even for hyperfine split levels at low magnetic fields, allows initialization of the spin state by cooling to experimentally attainable temperatures $\lesssim 10$ mK. Checking
 universality then reduces itself to showing that any gate operation connecting any two arbitrary
states can be realized within the spin decoherence time ${\rm T}_{2}$.\cite{Muthukrishnan2000,Brennen2005,Jenkins2017,Luis2020,Kraus2007}
The situation is illustrated in Fig. \ref{Universality:figure}-c and d for
a $d=8$ GdW$_{30}$ qudit. The plots show the rates $W_{n,m}$ of operations linking basis states $n$ and $m$, implemented by sequences of resonant transitions. The Rabi frequency $\Omega_{\rm R}$
sets the frequency uncertainty of a finite duration ($\sim 1/\Omega_{\rm R}$) pulse. Addressability can then be enforced by choosing only those transitions whose resonant frequencies
 fulfill $\omega_{j}-\omega_{i} > \Omega_{\rm R}$, for any $i$ and $j$. This level anharmonicity also allows reading out the spin states.\cite{Jenkins2016,Vincent2012,Godfrin2017b} A molecular spin qudit is universal when all $W_{n,m}T_{2} > 1$. The lowest $W_{n,m}T_{2}$
(Fig. \ref{Universality:figure} (e)) allows benchmarking the performances of different molecules or of molecules with respect to other schemes.\cite{Gimeno2021}  

For a given molecular qudit, $W_{n,m}$ depend on the relative strengths
of the Zeeman interaction, magnetic anisotropy and hyperfine couplings. Under carefully chosen conditions, it is possible to combine universal operation
with a large number of direct and fast links between pairs of states (compare panels (c) and (d) in Fig.
\ref{Universality:figure}). This possibility can reduce the number of operations required to implement certain gates and algorithms and thus help molecular NISQs to reach higher "quantum
volumes" \cite{Cross2019} than platforms based on linking nearest neighbour qubits. However, as discussed in Sec. IV, the number of levels within a molecular processor cannot be increased at will. At some point, actual scalability of the proposed platform requires to wire-up different molecular units by coupling them to resonant cavities.

\begin{figure*}[ht!]    
	\centering
	\includegraphics[width=0.9\textwidth]{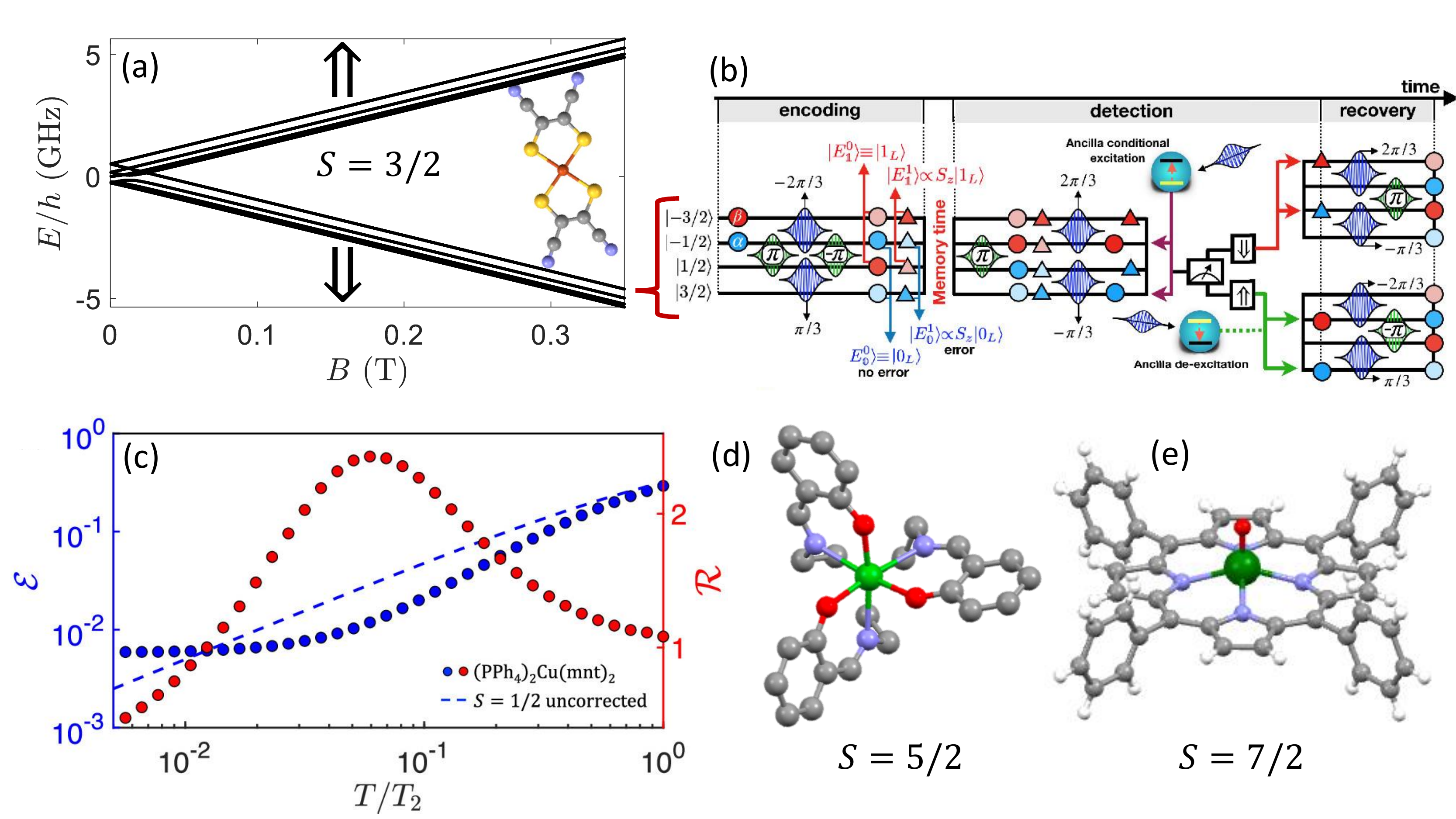}
	\caption{\label{QEC:fig} Molecular spin qubits with embedded QEC. (a) Level diagram of
	(PPh$_4$)$_2$[Cu(mnt)$_2$] (inset), embedding an $S=3/2$ nuclear spin
	\cite{Bader2014} hyperfine-coupled to an electronic spin 1/2, as a function of
	the external field, applied along $z$ axis. Energy levels are split by the
	electronic Zeeman interaction into a low-lying electronic $\Downarrow$
	subspace, in which the protected qubit is defined, and an excited electronic
	$\Uparrow$ manifold, with auxiliary levels needed for error detection. (b) Zoom
	on the electronic $\Downarrow$ subspace and corresponding sequence of
	radio-frequency/micro-wave pulses needed to correct an $S_z$ error on the
	qudit. Occurrence of this error brings each code word to an orthogonal state,
	thus making it possible to detect errors by a conditional excitation of the
	electronic ancilla.\cite{Chiesa2020} 
	(c) Simulated final error $\mathcal{E}$ (blue) as a function of the memory time $T$ in units
	of the qudit $T_2$ and relative gain $\mathcal{R}=\mathcal{E}_{1/2}/\mathcal{E}$ (red) with respect to an isolated spin $1/2$ (characterized by error $\mathcal{E}_{1/2}$) with no correction. Thus, $\mathcal{R}$  measures the reduction of error obtained by QEC.	
	(d,e) Other possible implementations using larger
	nuclear spin systems: Yb(trensal)\cite{Hussain2018} ($S=5/2$) and VOTTP
	($S=7/2$),\cite{Chicco2021} both coupled to (effective) electronic spin
	doublets. Reprinted (adapted) with permission from \onlinecite{Chiesa2020}. 
	Copyright American Chemical Society (2020).}

\end{figure*}

\section{\label{QEC:section} Quantum error correction and quantum simulations with molecular spin qudits}

The fundamental point behind {\it QEC} algorithms is to exploit a Hilbert space with
dimension larger than $2$ to encode a logical qubit. This extra space can make
errors detectable and correctable. \cite{Knill1997,Terhal2015} Furthermore, when
considering a specific physical implementation of qubits in the NISQ
era,\cite{Preskill2018} another important issue is to design QEC schemes correcting
the most important errors occurring in the real hardware. Here, the interaction
with neighboring nuclear spins leads to the decay of out-of-diagonal elements of
the density matrix of the central spins, thus destroying the quantum computation.
Hence, we focus on QEC codes designed to counteract dephasing and we discuss how
molecular nanomagnets can be exploited to define qubits with embedded QEC .\cite{Hussain2018,Chiesa2020,Macaluso2020} 

Two routes have been put forward to embed QEC in single magnetic molecules. On the
one hand, standard block codes can be efficiently implemented in molecules made of
weakly interacting spins $1/2$. \cite{Macaluso2020} On the other, effective qudit
QEC codes can be implemented exploiting the $d=2S+1$ levels of a spin $S$ system.\cite{Chiesa2020,Chiesa2021,Petiziol2021}

The first approach was investigated in Ref. \onlinecite{Macaluso2020}, where it was
shown that an [ErCeEr] trimer is a promising molecule to encode a logical qubit
protected against dephasing by the three-qubit phase-flip code. This
implementation requires three weakly interacting (effective) spin $1/2$, in order
to avoid the occurrence of correlated errors which are not handled by this code.
However, the interaction between the qubits must be sufficient to enable
excitations of one of the spin conditioned to the state of the others. Moreover,
significantly different $g$ values enable the use of fast manipulation pulses,
which are crucial because dephasing acts also during the implementation of QEC.
Rare-earth Kramers ions such as Er and Ce perfectly fit these
requirements.\cite{Macaluso2020}

The second approach was proposed in Ref. \onlinecite{Chiesa2020}. It exploits a
single electronic or nuclear spin qudit $S$ to encode an error-protected qubit. 
This is done by designing code words (consisting of superpositions of qudit states)
which are robust against error-operators characterizing the incoherent
dynamics of the central spin. Specialization to real
errors gives a substantially better performance
compared to abstract generic error models.\cite{Cafaro2012} In the simplest modelling of the bath, error operators are derived from a perturbative
expansion of the solution of the Lindblad equation,\cite{Chiesa2020} but the
derivation of optimized code words can be extended to more realistic nuclear bath
dynamics.\cite{Petiziol2021}

Once the code words have been determined, a proper sequence of electromagnetic
pulses can be designed to
implement the QEC code on a given molecular hardware (see Fig. \ref{QEC:fig}-(a,b)). The simplest physical
realization is represented by a
nuclear spin $S$ qudit coupled to a spin $1/2$ electronic ancilla, used to detect errors. 
The gain is remarkable already for a minimal $S=3/2$ qudit, realized e.g. in
(PPh$_4$)$_2$[Cu(mnt)$_2$] complex \cite{Bader2014} (Fig. \ref{QEC:fig}-(a)). It shows a maximum at intermediate $T/T_2$ (Fig. \ref{QEC:fig}-(c)), due to the finite
duration of the QEC procedure included in the
simulation, which bounds the error for small $T/T_2$.
Being effective also for long memory times, the scheme allows many gates to be
implemented before error
correction is needed, which could constitute a very
important advantage in the NISQ era.

Both approaches outlined above are able to defeat pure dephasing 
already in the two minimal implementations, represented by $3$ spins $1/2$ or by a
spin $3/2$ qudit \cite{Macaluso2020,Chiesa2020}.
The latter however appears simpler (a single spin) and
even easier to scale up. Indeed, the performance of the
code can be improved by increasing the number of qudit
levels, thus making it possible to correct higher-order
dephasing errors. This can be done by considering
larger nuclear spins, such as $^{173}$Yb in Yb(trensal) \cite{Hussain2018}
($S=5/2$) or $^{51}$V in VOTPP ($S=7/2$),\cite{Chicco2021}  shown in 
Fig. \ref{QEC:fig}- (d,e). Some remarks are necessary 
here: although chemically easy, the extension of the Hilbert space must be combined
with the design of suitable code words showing a large gain even at intermediate
times \cite{Petiziol2021} or of shorter pulse sequences whose duration does not
strongly increase with the number of levels. This requires that all addressed energy gaps are well separated in the spectrum (i.e. $\mid \omega_{i}-\omega_{j} \mid > \Omega_{\rm R}$ for significantly large $\Omega_{\rm R}$). This condition translates into
significant real or effective quadrupole interactions\cite{Chicco2021,Gimeno2021} (in the case of a nuclear spin qudit) or zero-field splitting (for electronic spin systems).\cite{Jenkins2017,Luis2020} Increasing the frequency separations allows decreasing
the duration of the control pulses, which is  
fundamental to reduce the harmful effect of decoherence during the correction protocol.

A {\it digital quantum simulator} is a device able to
efficiently 
mimic the dynamics of a quantum
system different from the hardware.\cite{Tacchino2019} This can be done by first
mapping the target Hamiltonian onto the hardware and then decomposing the
corresponding dynamics into a sequence of elementary
one- and two-qubit gates, controlled by the
experimenter, via the Suzuki-Trotter decomposition.\cite{Lloyd1996}

The simplest quantum simulator based on molecular nanomagnets consists of a
molecular chain of alternating spin 1/2 qubits and different magnetic units acting
as a switch for the qubit-qubit
interaction.\cite{Santini2011,Chiesa2014,Ferrando2016a,Ferrando2016b} 
The latter is effectively turned on by a conditional excitation of the switch,
depending on the state of
neighboring qubits via the qubit-switch coupling. This implements an entangling
two-qubit controlled-phase gate, which, combined with single-qubit rotations,
forms a universal set of gates and hence enables
digital quantum simulation of a wide class of models,
such as spin $\ge 1/2$ chains and fermionic systems.\cite{Santini2011,Atzori2018a,Chiesa2015} 

Very recently, this idea has been extended to include units with $S>1/2$.\cite{Tacchino2021} The multiple levels available within each qudit can
simplify quantum simulation of models involving several degrees of freedom, such as
bosonic fields interacting
with matter.\cite{Tacchino2021} The description of photon
modes (including in principle an infinite number of
levels) is a difficult task for qubit-based approaches,
yielding an exponentially large Hilbert space or
non-local interactions and thus deep quantum circuits.\cite{Sawaya2020,Mathis2020,DiPaolo2020} Both the number of
objects and the complexity of operations can be greatly simplified by pursuing a
qudit-based approach, in which
the photon space is truncated to the number of qudit levels.

The elementary unit of a molecular quantum simulator was realized in the following years,\cite{Chiesa2014,Ferrando2016a,Ferrando2016b} based on Cr$_7$Ni molecular qubits
with an interposed magnetic ion, in different geometries. This approach has however a possible
limitation: scaling up this platform by simply
extending the two-qubit units proposed in Refs. \onlinecite{Chiesa2014,Ferrando2016a,Ferrando2016b,Chiesa2016} can only be done up to about
ten qubits, due to residual qubit-qubit couplings which are not completely suppressed by
the scheme and become effective when elongating the register.
Nevertheless, this idea can be combined with photon-induced
coupling between different molecular processors to scale up
(see next section).

A nuclear spin implementation of a molecule-based digital quantum simulator was also
proposed, with the electronic spins mediating the coupling between qubits encoded in
the nuclear spins of $^{51}$V ions.\cite{Atzori2018a} This approach shows some remarkable
advantages: first, nuclear spins are characterized by remarkably longer coherence,
compared to their electronic counterpart. This benefit is usually
canceled by slow nuclear qubit operations. In the
scheme proposed in Ref. \onlinecite{Atzori2018a}, the use of
fast electronic excitations makes two-qubit operations
between nuclear spin qubits much faster than in
standard NMR approaches, thus exploiting the long nuclear coherence.
Second, large nuclear spins ($7/2$ in the case of $^{51}$V) paves the way to the use of
qudits, embedding quantum error correction. A scheme for implementing two-qubit gates on such
error protected nuclear spins has been recently put forward\cite{Chiesa2021} and it
could be extended in the near future to a more general class of gates and to quantum
simulations.

\section{\label{Wireup:section} Wiring-up molecular spin qudits}
While one could continue to increase the dimension of the molecular qudits,
the progress in computational power will eventually be limited by serious technical
difficulties. A major problem is associated with the "frequency crowding" of the required set of resonant transitions. This effect increasingly hinders addressing them spectroscopically, as it can already be seen in Fig. \ref{Universality:figure} (b) for $d=64$. There have proposals for introducing switchable couplers within molecular or supramolecular structures, formed by either ancillary spin
qubits\cite{Ferrando2016b} or by molecular linkers that can modify its electronic structure
under some external stimulus.\cite{Salinas2017} Yet, at some point, this path must be
complemented with the ability to locally control and, especially, wire up in a tunable manner different individual molecular spins.

\begin{figure*}
\includegraphics[width = 0.9\textwidth]{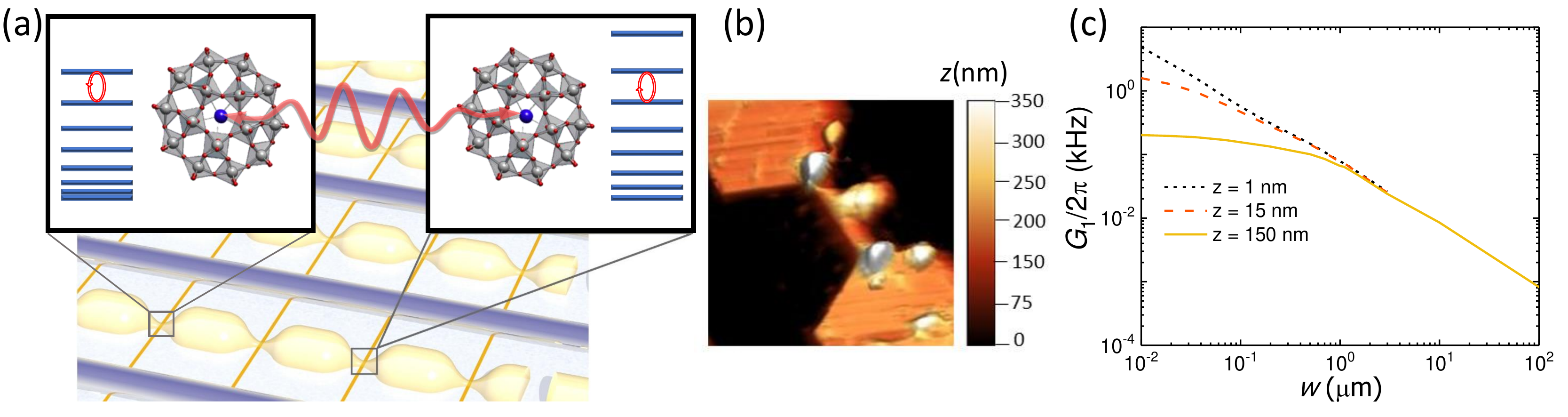}
\caption{\label{Wireup:figure} (a) Scheme for wiring up molecular spin qudits through their 
coupling to a common cavity mode in a superconducting resonator. The photon mediated 
interaction between suitably tuned qudits allows the implementation of two-qudit gates and 
ensures universal operation. This scheme allows different levels of scalability: using 
multiple levels inside each qudit, wiring up several of them in a resonator and coupling 
different resonators on a chip. (b) Atomic Force Microscopy (AFM) image of a nanoconstriction
fabricated in the central line of a superconducting resonator and of nanodrops of DPPH
$S=1/2$ free radical molecules deposited onto it by the same AFM tip. (c) Dependence of the 
single spin (for $S=1/2$) to single photon coupling as a function of the central line width 
$w$ and of the distance $z$ between the molecule and the constriction.}
\end{figure*} 

An advantage of molecular nanomagnets in connection with this idea is that most of them are
stable as individual units, e.g. in solution, and therefore can be transferred onto a solid
substrate or a device.\cite{Mannini2010,Domingo2012,Malavolti2018} This has enabled the 
realization of spin-dependent electron transport experiments on single molecules. These experiments allow reading out and coherently controlling their nuclear\cite{Vincent2012,Thiele2014} and
electronic spin states\cite{Godfrin2017b} and that have provided a first proof-of-principle
realization of the Grover quantum search algorithm with a nuclear spin
qudit.\cite{Godfrin2017a}

Here, we consider scaling up via the coupling of molecular spins to superconducting on-chip 
resonators.\cite{Jenkins2016} The basic idea is to adapt to the 
realm of molecules techniques of circuit quantum electrodynamics, which was originally
introduced as a platform for reading out and coherently
communicating superconducting qubits.\cite{Blais2004} The scheme is
shown in Fig. \ref{Wireup:figure}. A number $N$ of
magnetic molecules couple to the cavity mode of a
resonator. Their spin states and energy levels can
be controlled by a combination of global and local
magnetic fields. When the coupling energy $G$ of each
spin to a single photon is larger than the decoherence rates of the spin $1/T_{2}$ (typically $> 10^{4}$ Hz) and of the cavity $\kappa$ (typically $< 10^{4}$ Hz), the shifts it induces on the resonance frequency $\Omega$ can be determined from the
transmission through the device and it allows reading
out the spin state. More importantly for our purpose
here, it also introduces effective interactions between the spins. \cite{Zueco2009} We have 
generalized this effective photon-mediated interactions to the case of spin qudits having 
multiple levels.  The effective interaction Hamiltonian between two spin qudits is
\begin{equation}
{\cal H}_{J} =  \Omega \sum_{\vec{\alpha}, \vec{\beta} = 1}^{2S+1} \lambda_{1}^{\vec{\alpha}} \lambda_{2}^{\vec{\beta}} \left( \frac{1}{ E_{\vec{\beta}}^{2} - \Omega^{2}} + \frac{1}{ E_{\vec{\alpha}}^{2} - \Omega^{2}} \right) X_{1}^{\vec{\alpha}} X_{2}^{\vec{\beta}}
\label{Spin-photon:equation}
\end{equation}

\noindent where $\vec{\alpha} \equiv (\alpha_{1},\alpha_{2})$ denotes two eigenstates of one spin, separated by an energy gap $E_{\vec{\alpha}} \equiv E_{\alpha_{1}} - E_{\alpha_{2}}$
and connected by the Hubbard operator
$X_{1}^{\alpha_1,\alpha_2} \equiv \left| \alpha_{1} \right\rangle \left\langle \alpha_{2} \right|$ (and similarly for $\vec{\beta}$ in the other spin).
Constants $\lambda_{1}^{\vec{\alpha}}$ and $\lambda_{2}^{\vec{\beta}}$ depend on the wave
functions of these states and are proportional to the spin-photon coupling $G$.
Equation (\ref{Spin-photon:equation}) is derived in the dispersive regime, {\it i.e.} the spins' frequencies are non resonant with the cavity. Several architectures use such photon mediated couplings to generate two-qubit entangling gates \cite{Rigetti2010, Chow2011}. 
In particular, it allows to swap states of two qudits, as illustrated in Fig.
\ref{Wireup:figure}. Together with single-qudit operations reported in Fig. \ref{Universality:figure}, the coupling (\ref{Spin-photon:equation}) ensures complete control of the two-qudit system, thus forming a universal set.

This scheme works if a sufficiently strong coupling $G$ is attained. For
conventional coplanar superconducting resonators,\cite{Wallraff2004} the typical couplings amount to a few Hz, way below even the best decoherence
rates for molecular spins. In order to overcome this limitation, one needs to
bridge the gap that separates the sizes of the circuit and the molecule in order to locally enhance $G$. A direct method is decreasing the width of the
 inductor down to a few nm.\cite{Jenkins2013,Jenkins2014} It was predicted
theoretically,\cite{Jenkins2013,Jenkins2016} and
recently confirmed by experiments,\cite{Gimeno2020} that squeezing in
this manner the photon magnetic field can increase $G$ by several orders
of magnitude (see Fig.
\ref{Wireup:figure}). A complementary approach is to work with very low inductance $LC$ lumped-element resonators, which show large current densities at resonance.\cite{Sarabi2019} It is 
expected\cite{Gimeno2020} that the combination of both approaches can
take $G$ for single spins close to tenths of MHz. This would suffice to reach the coherent coupling regime for single molecular spin qudits provided that $T_{2} > 10 \mu$s, which seems feasible.

\section{\label{Outlook:section} Outlook and conclusions}

The previous sections show that artificial magnetic molecules can
contribute to reach higher levels of computational power along two
complementary directions: reducing the computational costs of some
algorithms and providing new methods for wiring up additional quantum
resources. A paradigmatic example is QEC. Encoding a
protected qubit in a single physical object can greatly
simplify the practical implementation of both error
correction and the quantum logic on the protected
subspace.\cite{Hu2019,Chiesa2021} In addition, the
codes are specifically adapted to the energy level
schemes of the molecular spin qudits and to their
dominant error sources. Even more, they can be optimized in several
ways. First, by carefully evaluating the interactions with
nuclear spins in the molecules and their effect of the
qudit states. Second, by optimizing the pulse sequences
used in the protocols. Similar considerations apply to
the implementation of some quantum simulations, which
benefit from the multiple level structure that is
inherent to the qudit and from the simplification associated with the
avoidance of non-local operations.

This approach has already led to a first proof-of-concept implementation of Grover's search
algorithm on a $d=3$ spin qudit.\cite{Godfrin2017a} We foresee that more
will follow in the next few years, increasing the complexity of the computational Hilbert 
space to $3-4$ qubits (or $d=8-16$), enough to realize the simplest
QEC codes and quantum simulations. An advantage of
several quantum simulation and computation algorithms realized
on a molecular architecture is that
they do not necessarily require measuring the response
of single molecules. Depending on the spin states involved, they can be performed by using
broad-band ESR or NMR experiments on magnetically diluted crystals of molecular qudits. For this reason, probably the best suited candidates are
qudits based on internal spin states of individual ions
(either electronic, nuclear or a combination of both). The main technical requisite is to attain sufficiently long $T_{2} > 10-20 \mu$s in these
crystals, which is still quite demanding but seems within reach for
sufficiently low spin concentrations and adequately designed ligands. A 
promising alternative is the use of on-chip resonators coupled to
transmission lines,\cite{Bienfait2016,Eichler2017,Probst2017,Bonizzoni2020,Gimeno2021} which can help to widen the frequency ranges for the
excitation and read-out as well as to attain stronger microwave magnetic
fields, thus faster operation rates.     

Reaching the next level in scalability necessarily involves a modular
approach and the ability of coherently exchanging information between
different spins. Progress along these lines will probably rely on
a combination of bottom-up strategies with solid-state
circuits fabricated by top-down lithography. We have
shown in section \ref{Wireup:section} that the coherent
coupling to on-chip superconducting resonators provides
a scalable path to wire up and perform universal
operations with several molecular spin qudits. In
principle, this route can give rise to processors
hosting several tens of qudits, each of them acting as a qubit with
embedded error protection. Besides, different resonators can be integrated in 
a chip using technologies developed for platforms based on superconducting
 circuits. Achieving the strong coupling of individual molecular processors
  with a resonator would also enable projective measurements of the molecular state. In contrast to NMR approaches, this makes it possible to initialize
   the register also by measuring. Moreover, it greatly enlarges the range of
    possible algorithms.

Yet, attaining $G T_{2} > 1$ for single spins is still
very challenging, even though it appears feasible in a
medium term. Enhancing the spin-photon coupling will
probably require the combination of optimally designed
resonators and a local confinement of photons near
superconducting nano-structures. Electric fields can be
confined much more easily than magnetic fields. Therefore, a way to improve on the latter aspect is to use molecules showing a large
spin-electric coupling.\cite{Liu2019} A different alternative is to exploit a 
resonant spin-photon coupling to exchange information between
spins.\cite{Carretta2013} Yet another one is to use the spin-magnon coupling.
The latter would play the role of microwave photons in the superconducting
resonators or cavities. Strong coupling between spins and the Kittel mode in
 a YIG nanosphere has been proposed.\cite{Neuman2020} The formalism advocated here can easily be exported to these new
architectures. Work on the devices must be accompanied by progress in the
deposition of few molecules with an exquisite control
over the position and the proper interface with the
circuit surface. An option is to profit from either
self-organization or synthesis of molecular arrays at
the surface.\cite{Urtizberea2020} From the molecular design, it is important
to maximize the transition matrix elements that
determine the value of $G$, e.g. by using low-lying
levels of high-spin molecules to define the qudit
states.\cite{Jenkins2013} 

We finally note that work along this direction has also a  
significant impact on 
very important applications beyond the realm of
quantum computing. Increasing the coupling of superconducting resonators
to spins contributes to the development of on-chip
magnetic resonance with a sensitivity able of detecting
samples of $\sim 10^{-2}$pico-L, or even magnetic excitations of individual 
magnetic nanostructures.\cite{Bienfait2016,Eichler2017,Probst2017,Gimeno2020}
The molecular approach has the
added value of serving as a suitable vehicle to deliver
diverse samples, as well as to improve their interface
with the circuit. Finally,  these systems are also ideal for exploring the
quantum electrodynamical control of matter \cite{Ruggenthaler2018} and, in particular, for modifying and controlling long-range magnetically ordered phases.\cite{roman2020}

In summary, even though unleashing the full potential
of this hybrid technology, i.e. creating large-scale
molecular spin-based processors, will likely require
further developments across different disciplines, the
demonstration of its key ingredients, namely the
operation over molecular scale NISQs and the ability to
coherently couple two of them, seem well within the
reach of current technologies. 

The data that support the findings of this study are available from the corresponding author upon reasonable request.
\begin{acknowledgments}
We thank J Román-Roche and  O. Roubeau for helping us with Figures 2 and 4 respectively. We acknowledge funding from the European Union’s
Horizon 2020 research and innovation programme (QUANTERA project SUMO, FET-OPEN grant 862893 FATMOLS), the Spanish MICINN (grants RTI2018-096075-B-C21, PCI2018-093116, PGC2018-094792-B-I00), the Gobierno de
Arag\'on (grant E09-17R-Q-MAD), the CSIC Quantum Technology Platform PT-001, and the Italian Ministry of Education and Research (MIUR) through the co-funding of SUMO.
\end{acknowledgments}


\bibliography{MolSpinscaling}

\end{document}